# Human Brain Mapping based on COLD Signal Hemodynamic Response and Electrical Neuroimaging

Revati Shriram
[1]Research Scholar, Sathyabama University, Chennai.
[2]Cummins College of Engg for Women, Pune, INDIA

M. Sundhararajan, PhD.
Shri Lakshmi Ammal Engg. College, Chennai, INDIA

Nivedita Daimiwal
Cummins College of Engg for Women, Pune, INDIA

## ABSTRACT

To understand 'Working of Human Brain', measurements related to the brain function are required. These measurements should be possibly non-invasive. Brain should be disturbed as less as possible during the measurement. Integration of various modalities plays a vital role in understanding the cognitive and the behavioral changes in the human brain. It is an important source of converging evidence about specific aspects of neural functions and dysfunctions under certain pathological conditions. Focal changes in cortical blood flow are tightly coupled with the changes in neuronal activity. This constitutes the option to map the hemodynamic response and infer principles of the cortical processing, even of complex tasks. The very high temporal resolution of EEG and good spatial resolution by NIRS make this concurrent measurement unique to study the spatio-temporal dynamics of large scale neuronal networks in the human brain. Such integration of two techniques will help to overcome the limitations of a specific method. Such as insensitivity of electroencephalogram (EEG) to unsynchronized neural events or lack of near infrared spectroscopy (NIRS) to low metabolic demand. A combination of EEG and NIRS will be more informative than the two separate analyses in both modalities.

## 1. INTRODUCTION

Non invasive exploration of human brain was always the topic of interest. Modern imaging methods provide the opportunity for non-invasive (i.e. *in vivo*) study of human organs and can provide measurements of local neuronal activity of the living human brain *(A Toga et al, 2001)*. These imaging modalities can be divided into two global categories: Functional Imaging or Structural Imaging *(D. Mantini et al, 2010)*. Functional imaging technique can be used along with the structural imaging to better examine the anatomy and functioning of particular areas of the brain in an individual. *(T Fox et al, 1994)*

## 1.1 Functional Imaging

Functional imaging represents a range of measurement techniques in which the aim is to extract quantitative information about physiological function from image-based data *(K Blinowska et al, 2009)*. The emphasis is on the extraction of physiological parameters rather than the visual interpretation of the images. Functional modalities include Single Positron Emission Computed Tomography (SPECT) and Positron Emission Tomography (PET), these are the nuclear medicine imaging modalities *(R. Misri et al, 2011)*. Along with them Functional Magnetic Resonance Imaging (fMRI). Electroencephalogram (EEG), Magnetoencephalogram (MEG), Electrical Impedance Tomography (EIT) can also be named as a functional imaging technique *(Nivedita et al, 2012)*.

## 1.2 Structural Imaging

Structural imaging represents a range of measurement techniques which can display anatomical information. These modalities include X-ray, Computer Tomography (CT), Magnetic Resonance Imaging (MRI), Transcranial Magnetic Stimulation (TCM) and Ultrasound (US). There are many reasons to determine the regional blood flow in organs such as in the brain or kidney, or in cancerous tissue regions of the body. For example, the assessment of cerebral blood flow and its autoregulation can be used to investigate the normal physiology and the nature of various diseases of the brain *(T. S. Koh, 2002)*. Also, the efficacy of radiotherapy treatment of cancer cells depends on the local oxygen concentration which is governed by the local blood flow. A convenient, minimally invasive method of assessing blood flow within organs is hence constantly being sought *(A Toga et al, 2001)*.

## 2. MATERIALS AND METHODS

Neuroimaing can be carried out by various methods such as Electroencephalogram (EEG), Functional Magnetic Resonance Imaging (fMRI), Near Infrared Spectroscopy (NIRS), Magnetoencephalogram (MEG), Transcranial Magnetic Stimulation (TCM) and Computed Tomography (CT).

## 2.1 Electrical Neuroimaging by EEG

EEG (electroencephalogram) reflects electrical activity of a multitude of neural populations in the brain. This signal is extremely complex, since EEG is generated as a superposition of different simultaneously acting dynamical systems. EEG signal originates mainly in the outer layer of the brain mainly known as the cerebral cortex, a 4–5mm thick highly folded brain region responsible for activities such as movement initiation, conscious awareness of sensation, language, and higher-order cognitive functions *(E.B.J. Coffey et al, 2010)*. EEG signal describes electrical activity of the brain measured by unpolarized electrodes and belongs to the group of stochastic (random) signals in frequency band of about 0 – 50 Hz with high time resolution *(T. Heinonen et al, 1999)*. In contrast, the anatomical localization of specific sources of the electrical activity is very imprecise. Electrical impulses, which come from deep centers of the brain, are not possible to measure directly using the scalp EEG approach *(R. Labounek* *el at, 2012*). EEG patterns are characterized by the frequency and amplitude of the electrical activity. Mu and Gamma waves are used in the development of BCI systems. Table 1 shows the various EEG waves and its frequency band.

**Table 1: EEG Frequency Bands (*M. Khachab et al, 2010*)**

| Wave (Frequency) | Function |
|---|---|
| Delta (0.5-4 Hz) | Are normal during drowsiness and early slow-wave sleep |
| Theta (4-7 Hz) | Arise from emotional stress, especially frustration or disappointment |
| Alpha (8-13Hz) | Alpha waves of moderate amplitude are typical of relaxed wakefulness and are most prominent over the parietal and occipital sites. |
| Beta (13-30 Hz) | Lower amplitude beta activity is more prominent in frontal areas and over other regions during intense mental activity. They are associated with an alert state of mind and can reach frequencies near 50 Hz during intense mental activity. |
| Mu (8-12 Hz) | Are linked to cortical motor activity and have been associated with beta activity. Mu waves diminish with movement or the intention to move. They occupy the same frequency band as Alpha waves. |
| Gamma (26-40 Hz) | Gamma waves are considered to reflect the mechanism of consciousness. |

The basic requirement of electrical neuroimging is the correct recording of the potential field on the scalp. The quality of the scalp potential map depends on technical aspects of the recording apparatus as well as correct positioning of the electrodes.

## 2.2 Hemodynamic Response by NIRS

Optical brain imaging exploit unique property of light to image the brain for clinical as well basic science related applications. Near Infrared (IR) light (wavelength 600 - 1000 nm) easily penetrates the biological tissue *(S. Erickson et al, 2009, M. Tamura et al, 1997)*. Figure 1 shows absorption spectra of HHB and HbO$_2$.

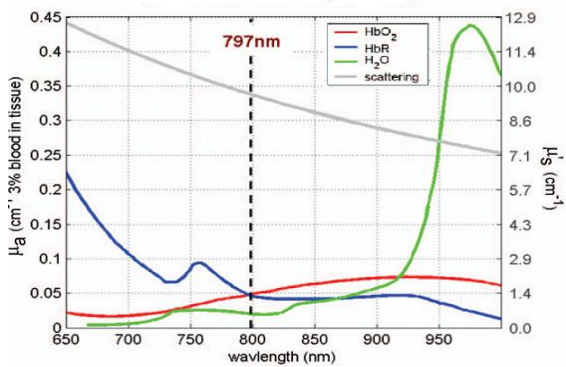

**Figure 1: Absorption Spectra of HHb and HbO$_2$ [22]**

NIRS is based on the observation that the properties of light passing through a living tissue are influenced by the functional state of the tissue, which is better explained mathematically by Beer Lamberts Law.

$$\frac{I}{I_O} = \exp(-\mu_a\, DPFx + G)$$

Where I is the detected intensity, I$_O$ is the incident intensity, DPFx is the mean path length travelled by the light in the tissue, G is a geometry dependent factor and $\mu_a$ is the absorption coefficient of the tissue *(E Hillman et al, 2007)*. Figure 2 shows the light propagation path inside the human skull.

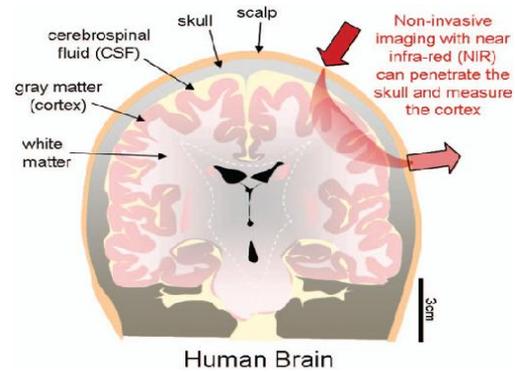

**Figure 2: Light Propagation Path inside the Skull [22]**

NIRS is a non-invasive method to measure oxygenation in a localized tissue and measures the transmission of infrared light through biological tissue *(G. Strangman et al, 2005)*. This indicates changes in oxygenation and the concentration of tissue chromophores such as total haemoglobin concentration (tHb) with its constituent oxygenated haemoglobin (HbO2) and deoxygenated haemoglobin (HHb) and cytochrome oxidase (CytOx) *(Nagdyman et al., 2003)*. NIRS signal is obtained based on capillary-oxygenation-level-dependent (COLD) signal. NIRS can asses two types of hemodynamic changes associated with the brain activity. Increase in neural activity results in increased glucose and oxygen consumption, which leads in increase in HbO$_2$ concentration. *(H. Matsuyama et al, 2009)*.

## 3. EEG-NIRS CONCURRENT MEASUREMENT

Integration is an important source of converging evidence about specific aspects and general principles of neural functions and dysfunctions in certain pathologies. Neuronal decoding w.r.t any behavioral movement or cognitive movement can be correlated by going for a concurrent measurement of EEG, MEG and NIRS *(M. Pflieger et al)*

### 3.1 System Setup

EEG-NIRS measurement depends on various physical properties such as conductivity, absorption and scattering coefficients of the head tissues such as scalp, skull, gray matter, white matter and Cerebral Blood Flow (CBF).

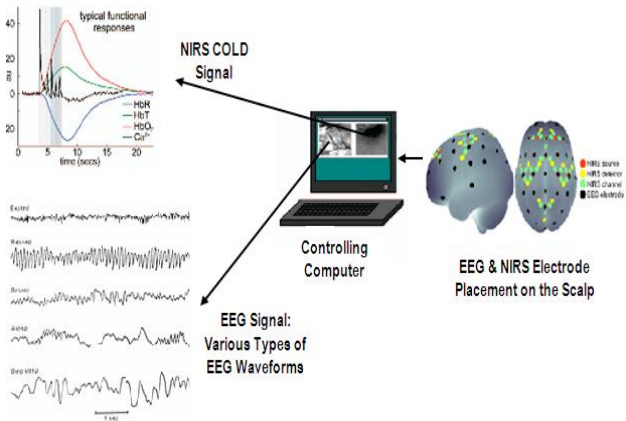

**Figure 3: EEG-NIRS System Setup**

NIRS requires the light in near infrared (NIR) region to determine cerebral oxygenation, blood flow and metabolic status of the brain. *(M. Pflieger et al)* It provides non-invasive means of monitoring the brain function and biological tissue because of relatively low absorption by water and high absorption by HHb and HbO2 in the range of 600-1000 nm wavelength *(Herve´ F et al, 2008)*. Figure 3 shows the system setup for EEG-NIRS concurrent measurement.

## 3.2 Interpretation

EEG electrodes and NIRS source-detector pair will be mounted on the human scalp, as shown in the figure 4. EEG signal will be captured while subject performing a cognitive/behavioral task. At the same time NIRS COLD signal will also be captured. COLD signal will reflect the change in HHb and $HbO_2$ with respect to the activity in the fronto temporal lobe. Increase in neural activity results in increased glucose and oxygen consumption, which leads in increase in $HbO_2$ concentration. Entropy based calculation can be carried for matching the electrophysiological response with the hemodynamic response. Figure 4 shows the illustrative flowchart of the EEG-NIRS concurrent measurement system. *F. Irani et al.* observed that when *e*ach time calculation was done, it caused a cognitive activation in the frontal region as demonstrated by increase in $HbO_2$ and decrease in HHb.

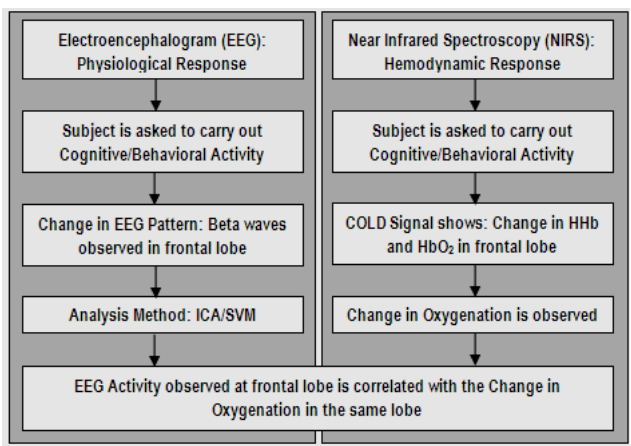

**Figure 4: Flowchart of EEG-NIRS Concurrent Measurement**

## 3.3 Data Analysis

EEG-NIRS data analysis is often made according to the spectral content of the recording signal. Following are the few methods used for this modality analysis.

- *WT: Wavelet Analysis*

WT is designed to address the problem of non stationary signals. It involves representing a time function in terms of simple fixed building blocks known as wavelets. Wavelets are ideally suited for sudden short duration signal changes. It has the ability to compute and manipulate data in compressed parameters which are known as features. So time varying signals such as EEG and NIRS consisting of many data points can be analyzed by using WT. WT can be characterized into continuous and discrete. Continuous wavelet transform (CWT) is given by

$$CWT(a,b) = \int_{-\infty}^{\infty} X(t)\varphi_{a,b}(t)dt$$

Where, X(t) is a analyzed signal, a & b represent the scaling factor and $\varphi_{a,b}$ is obtained by scaling the wavelet at time b & scale a *(D Cvetkovic et al, 2008)*.

- *ICA: Independent Component Analysis*

Recently developed ICA is an efficient tool for the artifact identification and extraction from the biosignals *(A. Hyvarinen et al, 2009)*. ICA transforms the experimental coordinate system into a non orthogonal basis that finds statistically independent components. It is used to solve the blind source separation problem to find linear coordinate system. In this case resultant signal will be statistically independent. ICA assumes that the observed data vector is a linear combination of unknown and statistically independent sources. The algorithm finds the demixing matrix W. *(Anita et al, 2007)* where

Data Vector = $\overline{X(t)} = [\overline{X_1(t)}, ...., \overline{X_n(t)}]^T$
Source Vector = $\overline{S(t)} = [\overline{S_1(t)}, ...., \overline{S(t)}]^T$ for T = 1,2,..P
$\overline{S(t)} = W\overline{X(t)}$

- *Spectral Analysis*

This analysis method is used to measure the coherence between the two electrodes. It asses the similarity of the spectral content of two electrodes over the time to reflect the measure of coupling between the various brain regions. *(A. Hyvarinen et al, 2009)* To extract the particular frequency feature based on the maximum amplitude, the data points are transferred to the frequency domain. This is performed by taking the FFT of the filtered data. The output of the FFT gives the power spectral values. *(S. Deivanayagi et al, 2007)*

- *SVM: Support Vector Machine*

This is one of the methods used for classification of neuroimging data. The classification task is based on the two separate sets of the training data and testing data. Each instance in the training set contains a target value and several attributes/features. In this case SVM predicts target values in the testing set when the features (decision boundary) are specified. In the formulation of SVM, the input vector $X_t$ is mapped to a high dimensional feature space $Z_t$ through a non linear transformation function ′g′. Where,

$$Z_t = g\,X_t$$
$$D\,Z_t = (W\,Z) + w_o$$

The SVM algorithm finds the decision boundary in the feature space given by the decision function. Where, W decides the linear decision boundaries. (*R. Sitaram et al, 2007*)

### 3.4 Applications

This concurrent modality has been used to investigate the synchronized activities of neurons and the subsequent hemodynamic response in human subjects. This simple and comparatively low-cost setup allows to measure hemodynamic activity in many situations when fMRI measurements are not feasible, (*A. Buchweitz et al, 2009*) e.g. for long-term monitoring at the bedside or even outside the lab via wireless transmission. (*H. Laufs et al, 2003*)

### 3.5 Limitations of EEG-NIRS Fusion

In this fusion, mapping of electrophysiological response with the hemodynamic response is discussed. Principal limitation on this multimodal analysis is imposed by the physiology of the brain itself. EEG and NIRS sources may be dislocated, distance between neuronal population whose electrical activity is recorded in EEG signal and the vascular tree which provides the blood supply to these neurons; since COLD signal recorded is a hemodynamic response. In addition to pre- and post-synaptic electrochemical dynamics, a number of physiological processes also require energetic support; for example, neurotransmitter synthesis, glial cell metabolism, maintenance of the steady-state transmembrane potential *(I. Kida et al, 2001)*, etc. These phenomena may cause hemodynamic COLD changes, without EEG correlates. This differential sensitivity to neuronal activity and energetic can also arise whenever hemodynamic activity is caused by non-synchronized electrophysiological activity (*Revati et al, 2012*). If the electrophysiological activity is transient, it might not induce any significant (i.e. detectable) metabolic activity changes *(Nunez et al, 2000, J. Daunizeau et al).*

## 4. DISCUSSION

Better temporal resolution with low spatial resolution is achieved by the EEG while good temporal resolution and moderate spatial resolution is offered by the NIRS (*P. Nunez et al, 2000*) Electroencephalography and Near Infrared Spectroscopy study simultaneously exploring electrophysiological and vascular response magnitude as a function of frequency during specific cognitive or behavioral activity. Changes in neuronal activity are tightly coupled to focal changes in cortical blood flow. This constitutes the option to map the hemodynamic response and infer principles of the cortical processing, even of complex tasks. Coupling between activity related electrophysiological and cerebrovascular changes allows high-resolution functional mapping. Larger decrease in HHb is interpreted as an indicator of an increase in blood flow velocity, it can only stem from an increased washout of HHb from the volume sampled.

## 5. CONCLUSION

NIRS can be applied in a variety of conditions as bedside monitoring in intensive care and in the operating theatre, where fMRI can be difficult to apply. All these non-invasive brain modalities complement and restrain each other and hence improve our understating of functional and neuronal organization. Spatially, temporally, physiologically, behaviorally and cognitively accurate computational models of the neuronal systems are the ultimate goals of the functional brain imaging. This goal can be achieved by integrating the diversity of various brain mapping techniques. By combining the various modalities together, we can exploit the strengths and flaws of individual brain imaging methods.

## 6. ACKNOWLEDGEMENT

The authors are grateful to Dr. Madhuri Khambete for her help towards the completion of this paper, as well as for providing valuable advice.

I would also like to thank my colleagues from Cummins College of Engineering for Women, for their feedback during the discussions.